# Calculation of The Critical Point for Two-Layer Ising and Potts Models Using Cellular Automata

Yazdan Asgari[1] , Mehrdad Ghaemi[2,3] , Mohammad Ghasem Mahjani[1]

[1] Department of Chemistry, K.N.Toosi University of Technology, P.O.Box 16315-1618, Tehran, Iran
yazdan1130@hotmail.com , mahjani@kntu.ac.ir
[2] Department of chemistry, Teacher Training University, Tehran, Iran
ghaemi@saba.tmu.ac.ir
[3] Atomic Energy Organization of Iran, Deputy in Nuclear Fuel Production, Tehran, Iran

**Abstract.** The critical points of the two-layer Ising and Potts models for square lattice have been calculated with high precision using probabilistic cellular automata (PCA) with Glauber alghorithm. The critical temperature is calculated for the isotropic and symmetric case ($K_x=K_y=K_z=K$), where $K_x$ and $K_y$ are the nearest-neighbor interactions within each layer in the $x$ and $y$ directions, respectively, and $K_z$ is the interlayer coupling. The obtained results are 0.310 and 0.726 for two-layer Ising and Potts models, respectively, that are in good agreement with the accurate values reported by others.

## Introduction

For many years, the lattice statistics has been the subject of intense research interests. Although, at zero magnetic field, there is an exact solution for the 2-dimensional (2-D) Ising model [1,2], however, there is no such a solution for the two-layer Ising and Potts models. The Potts models are the general extension of the Ising model with $q$-state spin lattice i.e., the Potts model with $q$ = 2 is equivalent to the Ising model. Although we do not know the exact solution of the two-dimensional Potts model at present time, a large amount of the numerical information has been accumulated for the critical properties of the various Potts models. For further information, see the excellent review written by Wu [3] or the references given by him.

The two-layer Ising model, as a simple generalization of the 2-D Ising model and also as an intermediate model between the 2-D and 3-D Ising models, has long been studied [4-7] and several approximation methods have been applied to this model [8-14]. In the absence of exact solution, simulation methods such as Monte Carlo [15] are most important tools for computation of the critical properties of the various Ising and Potts models.

In addition to the Monte Carlo method, it was proposed that the Cellular Automata (CA) could be a good candidate to simulate the Ising models [16]. In the last two decade a large amount of works were done for describing Ising models by the CA approach and a great number of papers and excellent reviews were published [17-22]. Most of the works that have been done until now are focused on the qualitative description of various Ising and Potts models or to introduce a faster algorithm. For example, the Q2R automaton as a fast algorithm was suggested which has been studied extensively [22-27]. It was so fast, because no random numbers must be generated at each step. But in the probabilistic CA, like Metropolis algorithm [28], generation of the random number causes to reduce the speed of calculation, even though it is more realistic for describing the Ising model.

In this article, at the first section we have used probabilistic CA with Glauber algorithm [29] for calculating the critical temperature of the two-layer symmetric Ising model on the square grid. We used a large lattice size (2500 × 2500) in simulation in order to reduce the finite size effects. It is shown that our results are in a good agreement with those obtained from the best numerical methods. In section 2 we have extended the method to the two-layer symmetric 3-state Potts model. The importance of this section is due to the fact that there is no well tabulated data for such a model.

## 1. Two-Layer Ising Model

Consider a two-layer square lattice with the periodic boundary condition, each layer with $p$ rows and $r$ columns. Each layer has then $r \times p$ sites and the number of the sites in the lattice is $2 \times r \times p = N$. We consider the next nearest neighbor interactions as well, so the number of neighbor for each site is 5. In the two-layer Ising model, for any site we define a spin variable $\sigma^{1(2)}(i,j) = \pm 1$ in such a way that $i = 1,...,r$ and $j = 1,...,p$ where superscript $1(2)$ denotes the layer number. We include the periodic boundary condition as

$$\sigma^{1(2)}(i+r, j) = \sigma^{1(2)}(i, j) \tag{1}$$

$$\sigma^{1(2)}(i, j+p) = \sigma^{1(2)}(i, j) \tag{2}$$

The configuration energy for this model may be defined as [14]

$$\frac{E(\sigma)}{kT} = -\sum_{i=1}^{r,*}\sum_{j=1}^{p,*}\sum_{n=1}^{2}\{K_x \sigma^n(i,j)\sigma^n(i+1,j) + K_y \sigma^n(i,j)\sigma^n(i,j+1)\} - K_z \sum_{i=1}^{r}\sum_{j=1}^{p}\sigma^1(i,j)\sigma^2(i,j) \tag{3}$$

where * indicates the periodic boundary conditions (eqs 1,2), and $K_x$ and $K_y$ are the nearest-neighbor interactions within each layer in the $x$ and $y$ directions, respectively, and $K_z$ is the interlayer coupling. Therefore, the configuration energy per spin is

$$e = \frac{E(\sigma)}{kTN} \tag{4}$$

The average magnetization of the lattice for this model can be defined as [15]

$$\langle M \rangle = \left\langle \sum_{i=1}^{r,*} \sum_{j=1}^{p,*} \sum_{n=1}^{2} \sigma^n(i,j) \right\rangle \tag{5}$$

and the average magnetization per spin is

$$\langle m \rangle = \frac{\langle M \rangle}{N} \tag{6}$$

The magnetic susceptibility per spin ( $\chi$ ) and specific heat per spin ($C$) is defined as [15]

$$\frac{\partial < M >}{\partial \beta} = \beta(< M^2 > - < M >^2) \tag{7}$$

$$\chi = \frac{\beta}{N}\left(\langle M^2 \rangle - \langle M \rangle^2\right) = \beta N\left(\langle m^2 \rangle - \langle m \rangle^2\right) \tag{8}$$

$$C = \frac{k\beta^2}{N}\left(\langle E^2 \rangle - \langle E \rangle^2\right) = k\beta^2 N\left(\langle e^2 \rangle - \langle e \rangle^2\right) \tag{9}$$

where $\beta = \frac{1}{kT}$.

## 1.1 Method

In the present work, we considered the isotropic ferromagnetic and symmetric case i.e. $K_x=K_y=K_z=K \geq 0$ . We have used a two-layer square lattice with $2500 \times 2500$ sites in each layer with the periodic boundary condition. The Glauber method [29] was used with checkerboard approach to update sites. For this purpose the surfaces of two layers are checkered same as each others. For updating the lattice, we use following procedure: after updating the first layer, the second layer could be updated. The updating of the spins is based on the probabilistic rules. The probability that the spin of one site will be up ( $p_i^+$ ) is calculated from [30]

$$p_i^+ = \frac{e^{-\beta E_i^+}}{e^{-\beta E_i^+} + e^{-\beta E_i^-}} \tag{10}$$

where

$$E_i^{\pm} = -K\{\sigma^n(i,j)\sigma^n(i+1,j) + \sigma^n(i,j)\sigma^n(i-1,j) + \sigma^n(i,j)\sigma^n(i,j+1) + \sigma^n(i,j)\sigma^n(i,j-1) + \sigma^n(i,j)\sigma^{n'}(i,j)\} \tag{11}$$

and

$$\sigma^n(i,j) = +1 \text{ for } E_i^+$$
$$\sigma^n(i,j) = -1 \text{ for } E_i^- \qquad (12)$$

and $\sigma^{n'}(i,j)$ is the neighboring site (*i*,*j*) in the other layer. Hence, the probability that the spin to be down is

$$p_i^- = 1 - p_i^+ \qquad (13)$$

The approach is as follow: first a random number is generated. If it is less than $p_i^+$, the spin of the site (*i*,*j*) is up, otherwise (it means that random number is greater than $p_i^+$), it will be down.

When we start CA with the homogeneous initial state (namely, all sites have spin up or +1), before the critical point ($K_c$), the magnetization per spin (*m*) will decay rapidly to zero and fluctuate around it. After the critical point, *m* will approach to a nonzero point and fluctuate around it; and with increasing of *K*, the magnetization per spin will increase. But at the critical point, *m* will decay very slowly to the zero point and the fluctuation of the system will reach to a maximum. For each *K*, the time that *m* reaches to the special point and starts to fluctuate around it is called the relaxation time ($\tau$). On the other words, the relaxation time is the time that the system is thermalized. The value of $\tau$ can be obtained from the graph of *m* vs. *t* (Fig. 1). One can see from these graphs that the relaxation time increases before critical point and reaches to a maximum at $K_c$, but after the critical point, $\tau$ decreases rapidly. So, in the critical point, the system last a long time to stabilized. Hence, the critical point may be obtained from the graph of $\tau$ vs. *K* (Fig. 2). The obtained critical point from this graph is 0.310 for the two-layer Ising model.

In our approach, we have calculated the thermodynamic quantities after thermalization of the lattice. In other words, first we let the system reaches to a stable state after some time step ($t=\tau$), and then to be updated up to the end of the automata ($t$=50000). For example to calculate the average value of magnetization per spin ($<m>$), one should add all values of *m* from the relaxation time up to the end of the automata (or end of the time step) and divide the result to number of steps. The other way for calculation of the critical point is the usage of $<m>$. By drawing the graph of $<m>$ vs. *K*, we may also obtain $K_c$. Fig. 3 shows the results of such calculation. As it is seen, before critical point ($K<K_c$), $<m>=0$ and after that ($K> K_c$), $<m>\neq 0$. The obtained values of the critical point from this approach is $K_c$ =0.310 for the two-layer Ising model.

For calculation of $\chi$ for each *K*, first we have calculated the value of $(m - <m>)^2$ in each time step. Then these values are averaged in a some way explained above. According to eq. 8 this average could be used for computation of $\chi$ . Using eq. 9 for calculation of the specific heat (C), we have done it in a same way described above. Figures 4 and 5 show the graphs of $\chi$ vs. *K* and *C* vs. *K*, respectively, for the two-layer Ising model. These graphs are the other ways for obtaining the critical point. The maximum of these graphs indicates the critical point. The obtained value for $K_c$ from these graphs is 0.310 for the two-layer Ising model.

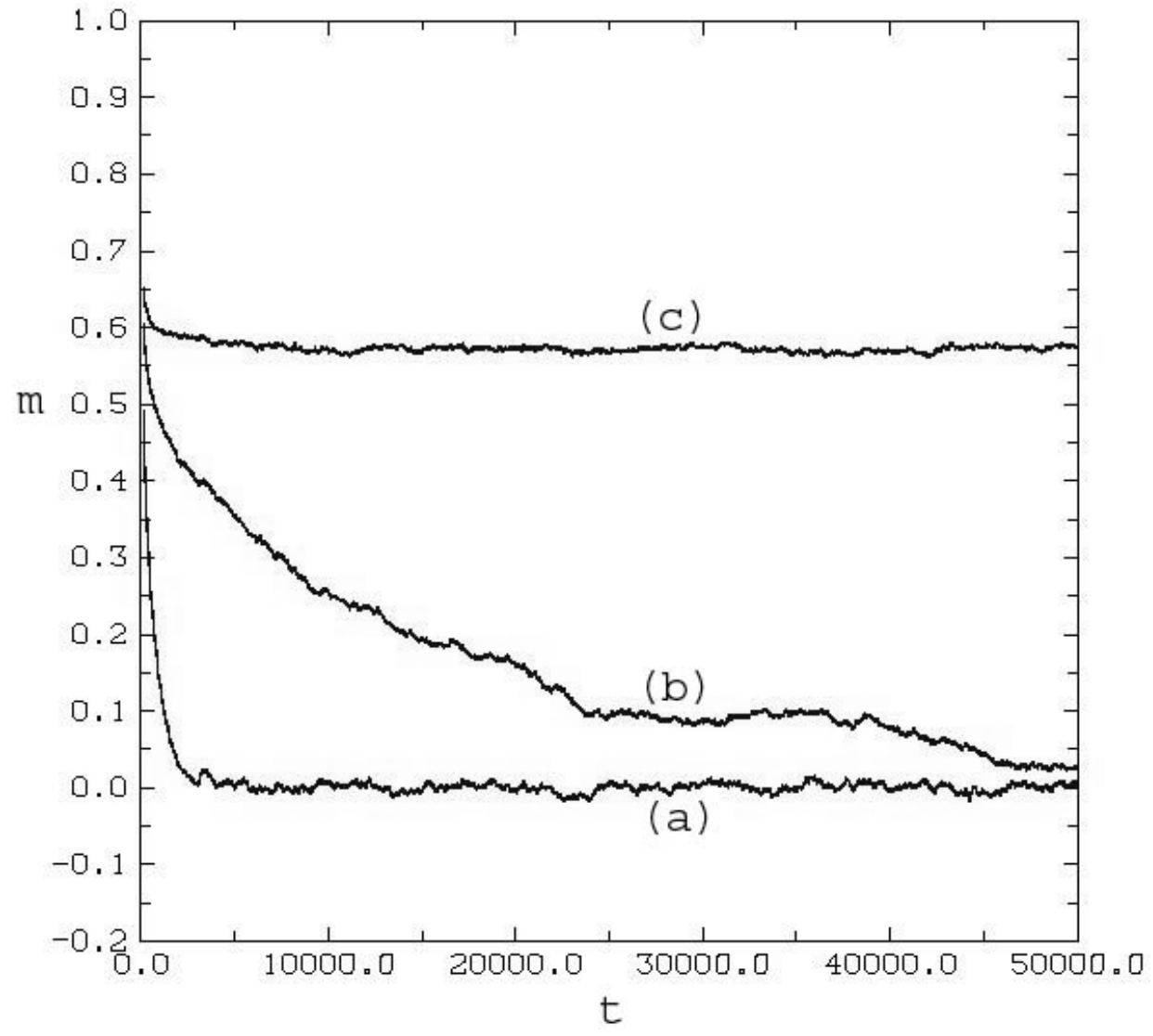

**Fig. 1.** The magnetization versus time in the two-layer Ising model. for 3 states. a: $K$=0.304 ($K<K_c$), $\tau$ =3500. b: $K$=0.310 ($K=K_c$), $\tau$ =46000. c: $K$=0.313 ($K>K_c$), $\tau$ =4000. (each layer has 2500$\times$2500 sites, start from homogeneous initial state "all +1", time steps = 50000)

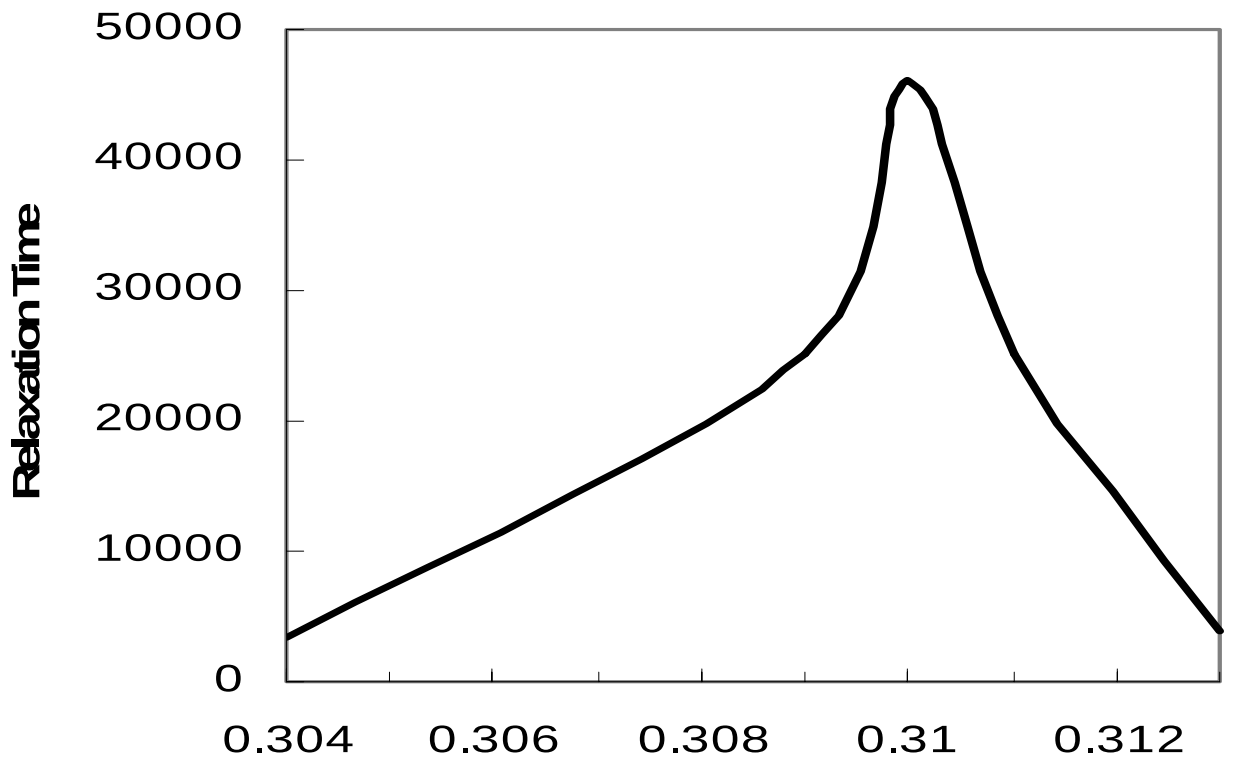

**Fig. 2.** The relaxation time obtained from Figure 1 versus $K$ for the two-layer Ising model. The maximum appears at $K=K_c$

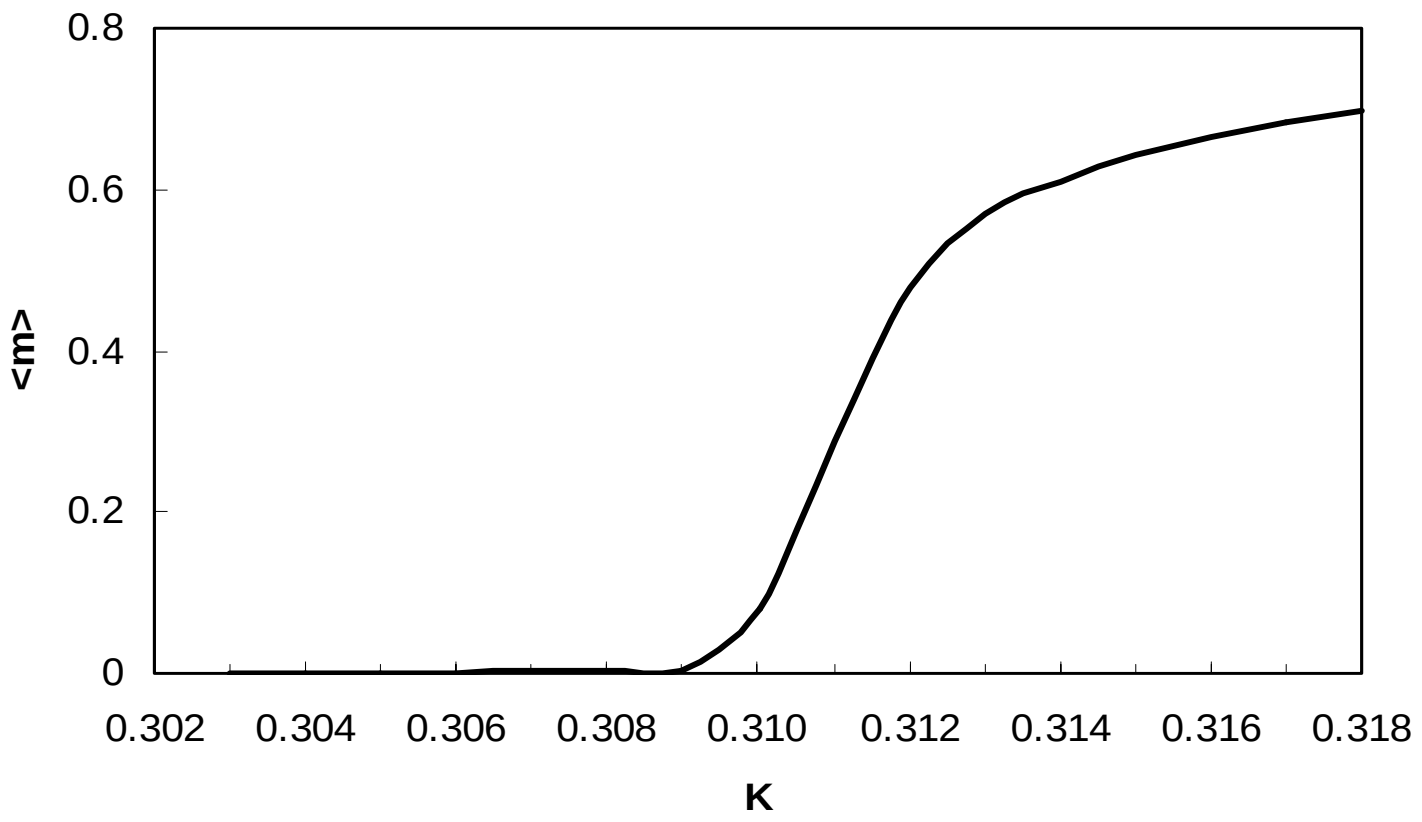

**Fig. 3.** <*m*> versus coupling coefficient (*K*) for the two-layer Ising model. The average value for each *K* is calculated after its relaxation time. (data are the results for the lattice that each layer has 2500 × 2500 sites, starting from the homogeneous initial state with all +1, time steps = 50000)

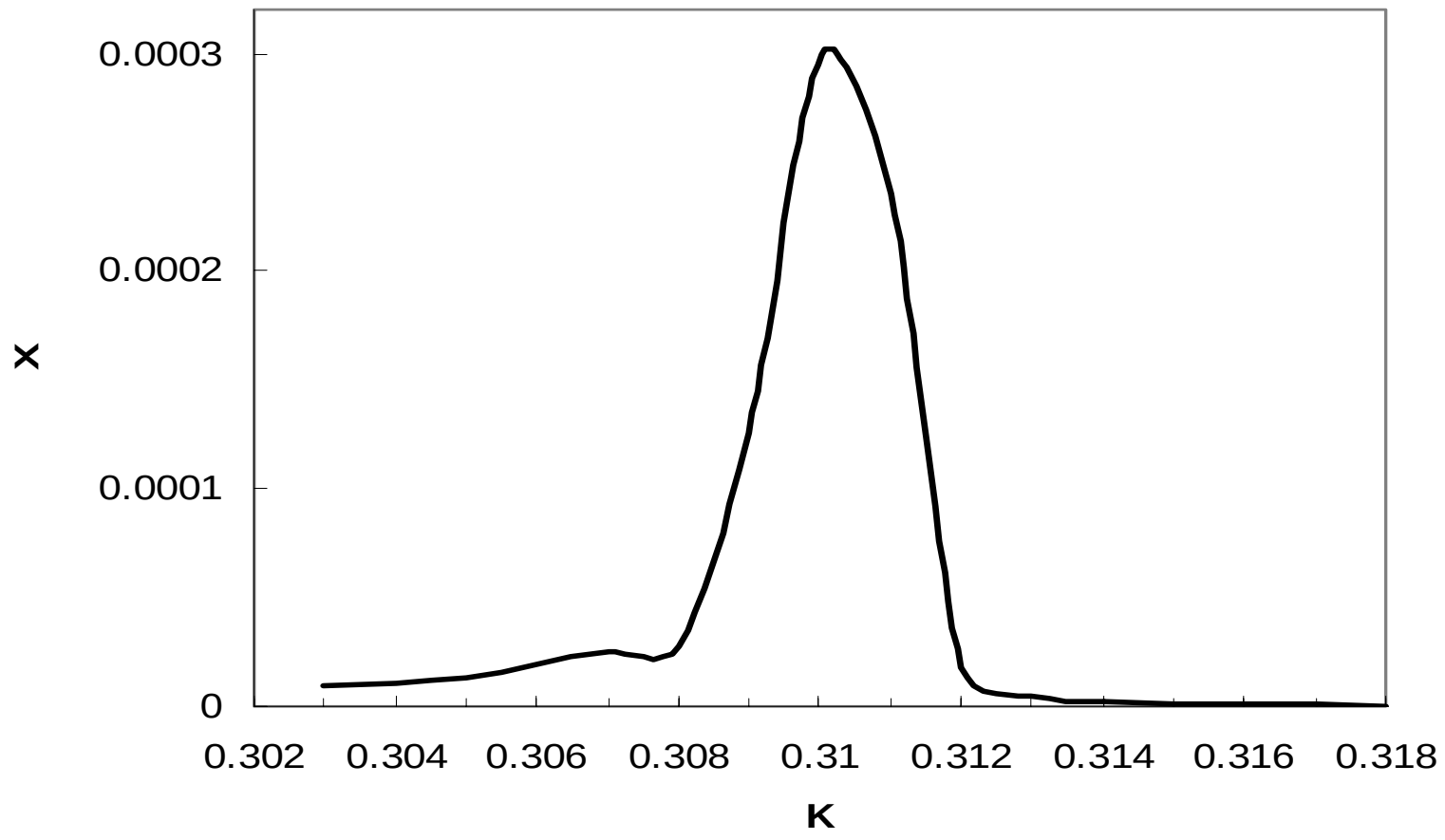

**Fig. 4.** Magnetization susceptibility per spin ( $\chi$ ) versus *K* for the two-layer Ising model. (The calculated data are the results for the lattice for which each layer has 2500 × 2500 sites, starting from the homogeneous initial state with all spins up, time steps = 50000)

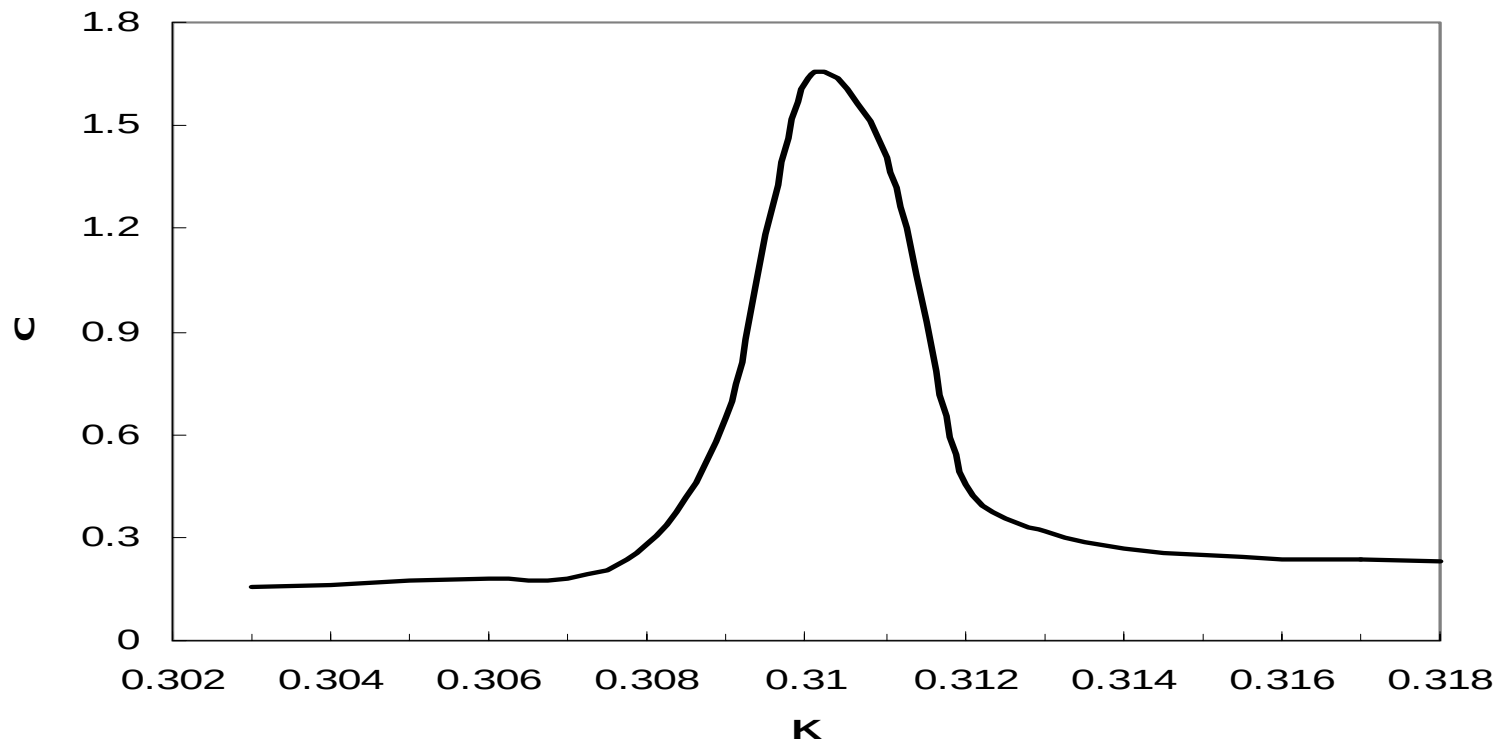

**Fig. 5.** Specific Heat per spin (*C*) versus *K* for the two-layer Ising model. (The calculated data are the results for the lattice for which each layer has $2500\times 2500$ sites, starting from the homogeneous initial state with all spins up, time steps = 50000)

## 2. Two-layer Potts Model

Consider a two-layer square lattice with the periodic boundary condition, each layer with *p* rows and *r* columns. Each layer has then $r\times p$ sites and number of sites in the lattice is $2\times r\times p = N$. We consider the next nearest neighbor interactions as well, so the number of neighbor for each site is 5. For any site we define a spin variable $\sigma^{1(2)}(i,j) = 0,\pm 1$ so that $i = 1,...r$ and $j = 1,..., p$. The configuration energy of the standard 3-state Potts model is given by [3],

$$\frac{E(\sigma)}{kT} = \sum_{i=1}^{r,*}\sum_{j=1}^{p,*}\sum_{n=1}^{2} -\{K_x\delta_{\sigma^n(i,j),\sigma^n(i+1,j)} + K_y\delta_{\sigma^n(i,j),\sigma^n(i,j+1)} + K_z\delta_{\sigma^1(i,j),\sigma^2(i,j)}\} \quad (14)$$

where

$$\delta_{i,j} = 1 \text{ for } i = j$$
$$\delta_{i,j} = 0 \text{ for } i \neq j \quad (15)$$

Other quantities are obtained from the equations 4-9.

### 2.1 Method

For quantitative computation of the two-layer Potts model, we have considered the isotropic ferromagnetic and symmetric case which $K_x=K_y=K_z=K\geq 0$. We have used a two-layer square lattice that each layer has $1500\times 1500$ sites with periodic boundary condition. Each site can have a value of the spin up (+1), down (-1) or zero (0). We used the Glauber method with checkerboard approach similar to the Ising model for updating the sites of the 3-state Potts model. The probability that the spin of one site will be up ( $p_i^+$ ) is calculated from

$$p_i^+ = \frac{e^{-\beta E_i^+}}{e^{-\beta E_i^+} + e^{-\beta E_i^-} + e^{-\beta E_i^0}} \tag{16}$$

and, the probability that the spin to be down is

$$p_i^- = \frac{e^{-\beta E_i^-}}{e^{-\beta E_i^+} + e^{-\beta E_i^-} + e^{-\beta E_i^0}} \tag{17}$$

Hence, the probability that the spin to be in zero state is

$$p_i^0 = 1 - (p_i^+ + p_i^-) \tag{18}$$

where

$$E_i^{\pm,0} = -K_x\{\delta_{\sigma^n(i,j),\sigma^n(i+1,j)} + \delta_{\sigma^n(i,j),\sigma^n(i-1,j)}\} - K_y\{\delta_{\sigma^n(i,j),\sigma^n(i,j+1)} + \delta_{\sigma^n(i,j),\sigma^n(i,j-1)}\}$$
$$- K_z\{\delta_{\sigma^n(i,j),\sigma^{n'}(i,j)}\} \tag{19}$$

and

$$\sigma^n(i,j) = +1 \text{ for } E_i^+$$
$$\sigma^n(i,j) = -1 \text{ for } E_i^-$$
$$\sigma^n(i,j) = 0 \text{ for } E_i^0 \tag{20}$$

The calculation steps are similar to the two-layer Ising model. Figure 6 shows the graph of <*m*> vs. *K* for the two-layer Potts model. Figure 7 and 8 are the graphs for $\chi$ and *C* vs. *K*, respectively. The obtained value of $K_c$ for the two-layer Potts model is 0.726. It is obvious that the value obtained from these figures, are in agreement.

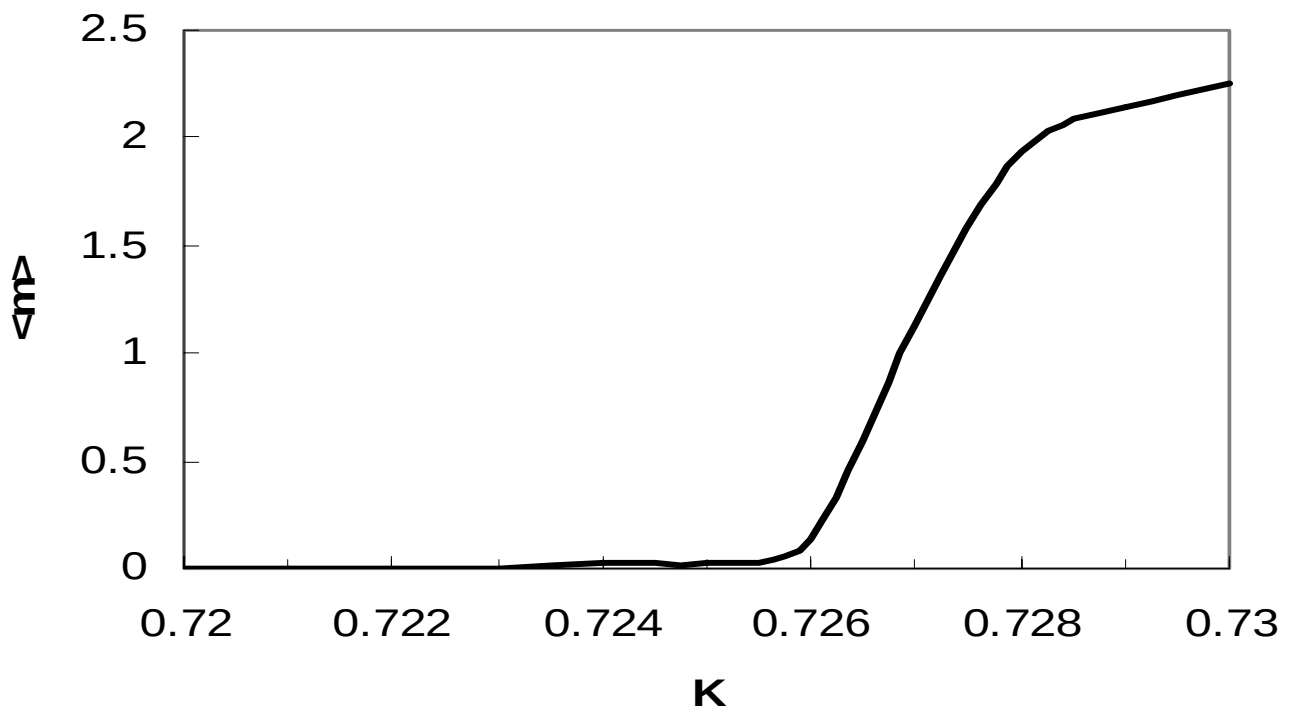

**Fig. 6.** <*m*> versus coupling coefficient (*K*) for the two-layer Potts model. (The calculated data are for the lattice that each layer has 1500×1500 sites, starting from the homogeneous initial state with all spins up, time steps = 50000)

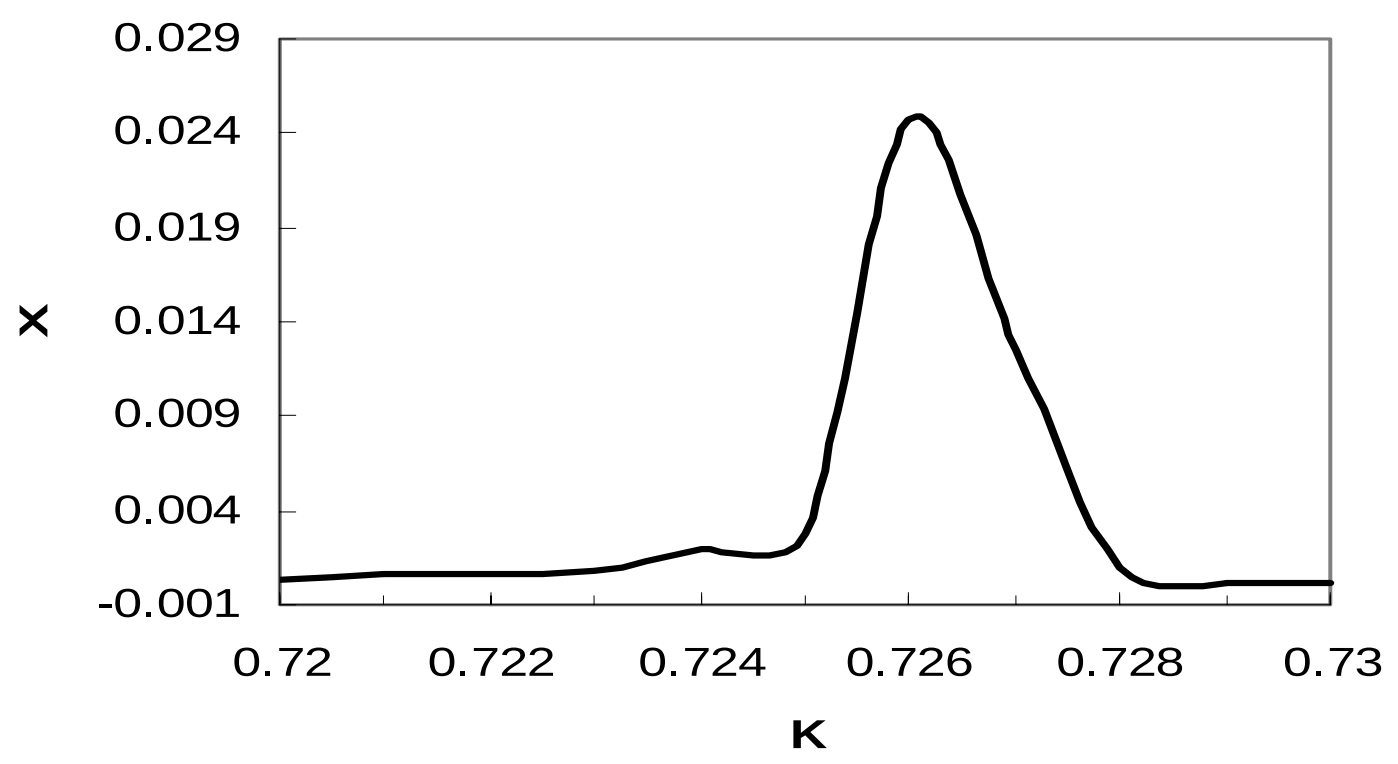

**Fig. 7.** Magnetization susceptibility per spin ($\chi$) versus $K$ for the two-layer Potts model. (The calculated data are for the lattice that each layer has $1500\times1500$ sites, starting from the homogeneous initial state with all spins up, time steps = 50000)

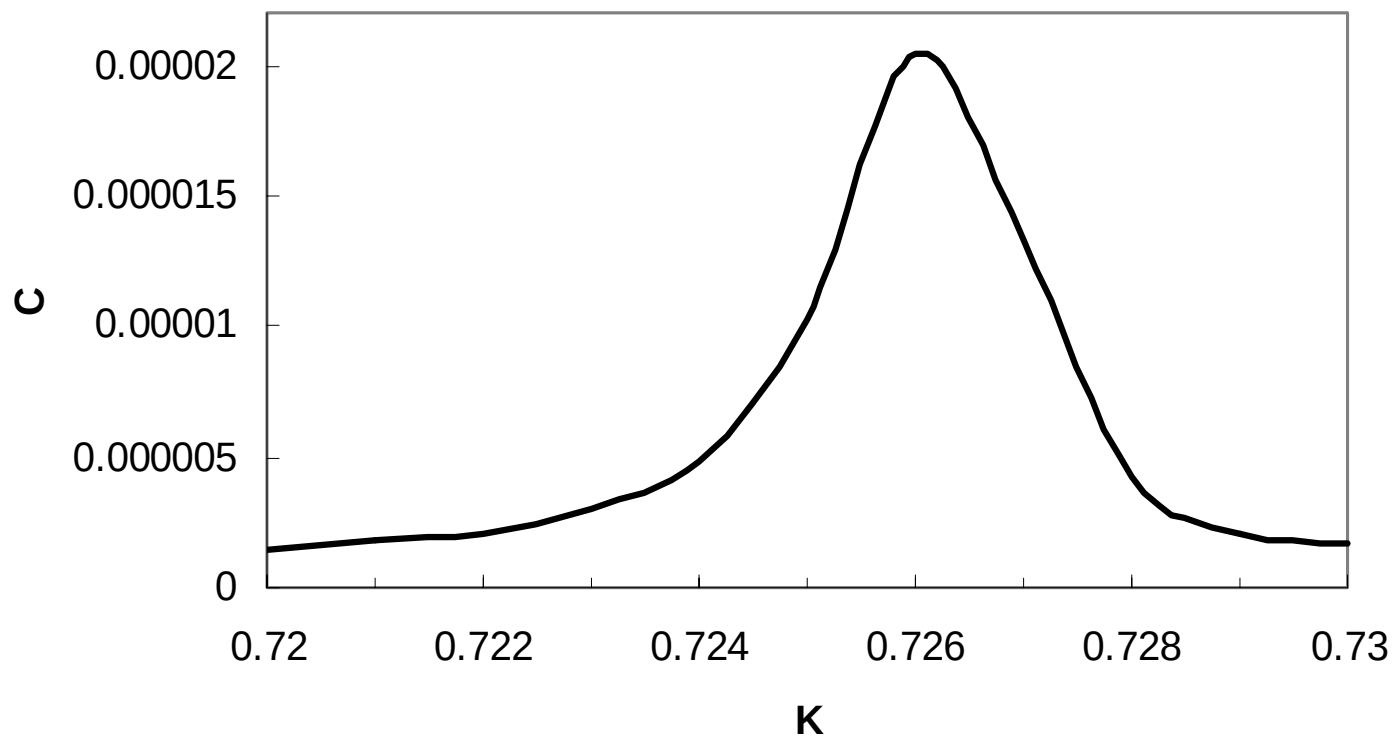

**Fig. 8.** Specific Heat per spin ($C$) versus $K$ for the two-layer Potts model. (The calculated data are for the lattice that each layer has $1500\times1500$ sites, starting from the homogeneous initial state with all spins up, time steps = 50000)

## Conclusion

It was demonstrated that the high precision calculation of the critical point can be done by the CA. For obtaining the fourth and further digits after the decimal point by the CA, one must use a larger lattice and since the relaxation time is large near the critical point, the numbers of time steps must be increased. For example, in order to compute the fourth digit of $K_c$ in the two-layer Ising model, it is sufficient to increase the number of time step up to 300000 steps and draw the graph <$m$> vs. $K$ . The calculated $K_c$ is 0.3108 that is in good agreement with other numerical method [14].

Although the extension of numerical methods for calculation of the critical properties of the two-layer 3-state Potts model is a difficult task, but such extension is easy in cellular automata approach. The importance of this approach is due to the fact that there is no well tabulated data for the two-layer Potts model.

## .Acknowledgment

We acknowledge Prof. G. A. Parsafar for his useful comment.